\PassOptionsToPackage{table}{xcolor}
\documentclass[runningheads]{llncs}
\usepackage{tikz}
\usetikzlibrary{fit,automata,positioning,calc}
\usepackage{algorithmicx,algpseudocode}
\usepackage{graphicx}

\usepackage{amsmath}
\usepackage{amssymb}
\usepackage{url}
\usepackage{multicol}

\newcommand{\HIDE}[1]{}

\newcommand{\transition}[3]{#1 {\stackrel{#2}{\longrightarrow}} #3}
\newcommand{\shorttransition}[2]{#1 {\rightarrow} #2}

\newcommand{\ON}{\; | \;}

\newcommand{\WORD}[1]{\langle #1 \rangle}

\newtheorem{mydef}{Definition}

\begin{document}
    
\title{Test Model Coverage Analysis under Uncertainty}

\author{I. S. W. B. Prasetya\inst{1}\orcidID{0000−0002−3421−4635}
        \and Rick Klomp\inst{1} 
       }

\institute{Utrecht University, the Netherlands, \email{s.w.b.prasetya@uu.nl} 
          }

\maketitle

\begin{abstract}
In model-based testing (MBT) we may have to deal with a non-deterministic model,
e.g. because abstraction was applied, or because the software under test itself is 
non-deterministic. The same test case may then trigger multiple possible execution paths,
depending on some internal decisions made by the software. 
Consequently, performing precise test analyses, e.g. to calculate the test coverage, are not possible.
This can be mitigated if developers can annotate the model with estimated probabilities
for taking each transition. A probabilistic model checking algorithm can subsequently
be used to do simple probabilistic coverage analysis. However, in practice developers 
often want to know what the achieved aggregate coverage, which unfortunately cannot be
re-expressed as a standard model checking problem. This paper presents an extension to allow efficient
calculation of probabilistic aggregate coverage, and moreover also in combination with 
$k$-wise coverage.

\keywords{probabilistic model based testing, 
          probabilistic test coverage, 
          testing non-deterministic systems}
\end{abstract}

{\scriptsize
\noindent
This is a {\bf preprint}. The final version of this paper is published in the Proceedings
of the International Conference on Software Engineering and Formal Methods
SEFM 2019, \url{https://doi.org/10.1007/978-3-030-30446-1_12}.
}

\section{Introduction}

\vspace{-3mm}
Model based testing (MBT) is considered as one of the leading technologies for systematic testing of software
\cite{PaleskaMBT2013,bringmann2008model,craggs2003agedis}. It has been used to test different kinds of software, e.g. communication protocols,
web applications, and automotive control systems. In this approach, a model describing the intended
behavior of the system under test (SUT) is first constructed \cite{utting2012taxonomy}, and then used to guide the tester, or a testing algorithm, to systematically explore and test the SUT's states.
Various automated MBT tools are available, e.g. JTorX \cite{JtroxPhdthesis,tretmans2003torx},
Phact \cite{heerink2000formal}, 
OSMO \cite{kanstren2012using},
APSL \cite{tervoort2017apsl},
and RT-Tester \cite{PaleskaMBT2013}.

There are situations where we end up with a non-deterministic model \cite{PaleskaMBT2013,stoelinga2009interpreting,jard2005tgv}, for example when
the non-determinism within the system under test,
e.g. due to internal concurrency, interactions with an uncontrollable environment
(e.g. as in cyber physical systems), or use of AI, leads to observable effects at the model level. 
Non-determinism can also be introduced as byproduct when we
apply abstraction on an otherwise too large model \cite{pretschner200510}. 
Models mined from executions logs \cite{ZellerMiningModels2013,ADABU2006,vos2014fittest}
can also be non-deterministic, because 
log files only provide very limited information about a system's states.

MBT with a non-deterministic model is more challenging. The tester cannot
fully control how the SUT would traverse the model, 
and cannot thus precisely determine the current state of the SUT. 
Obviously, this makes the task deciding which trigger to send next to the SUT harder.
Additionally, coverage, e.g. in terms of which states in the model have been visited by a
series of tests, cannot be determined with 100\% certainty either. This paper will focus
on addressing the latter problem ---readers interested in test cases generation from
non-deterministic models are referred to e.g. \cite{tretmans1992formal,nachmanson2004optimal,jard2005tgv}. 
Rather than just saying that a test sequence {\em may} cover some given state, we propose
to {\em calculate the probability} of covering a given coverage
goal, given modelers' estimation on
the local probability of each non-deterministic choice in a model.

Given a probabilistic model of the SUT, e.g. in the form of a Markov Decision Process (MDP) \cite{baier2008principles,stoelinga2002introduction},
and a test $\sigma$ in the form of a sequence of interactions on the SUT, the most elementary
type of coverage goal in MBT is for $\sigma$ is to cover some given state $s$ of interest in the model.
Calculating the probability that this actually happens is an instance of the probabilistic reachability problem which 
can be answered using e.g. a probabilistic model checker
\cite{hansson1994logic,baier2008principles,kwiatkowska2007stochastic}.
However, in practice coverage goals are typically formulated in an 'aggregate' form, e.g. to cover at least 80\% of
the states, without being selective on which states to include. Additionally, we may want to know the aggregate coverage over pairs of states (the transitions
in the LTS), or vectors of states, as in $k$-wise coverage \cite{ammann2016introduction}, as
different research showed that $k$-wise
greatly increases the fault finding potential of a test suite \cite{petke2015practical,grindal2005combination}.
Aggregate goals cannot be expressed in LTL or CTL, which are the typical formalisms in model checking.
Furthermore, both types of goals (aggregate and $k$-wise) may lead to combinatorial explosion.

This paper {\bf contributes}: (1)
a concept and definition of probabilistic test coverage; as far as we know this has not been covered in the literature before,
and (2)
%
 %
 an algorithm to calculate probabilistic coverage, in particular of aggregate $k$-wise coverage goals.

\HIDE{
Our contributions are as follows:

\begin{enumerate}
    \item We propose a concept of probabilistic test coverage. 
    Subsequently we propose a simple language to express coverage query, e.g. to
    calculate the probability that a test suite covers a certain state, or the
    probability that the test suite covers at least $N$ states. The latter kind
    of query is called {\em aggregate query}. The language is a subset
    of Probabilistic CTL (PCTL) \cite{baier2008principles}, but it also
    extends PCTL with aggregate queries.
    
    \item Non-determinism can explode the number of execution paths to consider when analyzing a test 
    suite. If done naively coverage analyses/calculation will explode as well.    
    We will show an algorithm that in most cases will perform efficiently. The corner case will
    be discussed, along with a recommendation for future research.
    Our algorithm can be seen as an extension of PCTL model checking (for the subset
    that we include) with the calculation of aggregate queries.

    This paper also includes benchmark results of the algorithm.
    
\end{enumerate}
}

{\bf Paper structure.} Section \ref{sec.probmodel} introduces 
relevant basic concepts.
%
Section \ref{sec.cov} introduces the kind of coverage goals we want to
be able to express and how their probabilistic coverage can be calculated.
Section \ref{sec.efficient.cov.calc} presents our algorithm for efficient coverage calculation.
Section \ref{sec.experiment} shows the results of our benchmarking.
Related work is discussed in Section \ref{sec.relatedwork}.
Section \ref{sec.concl} concludes.

\section{Preliminary: Probabilistic Models and Simple Coverage} \label{sec.probmodel}

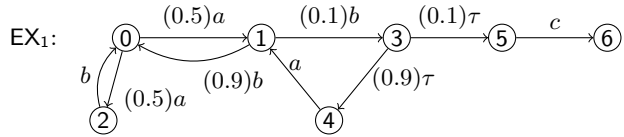
\begin{figure}
%
\begin{center}
\begin{tikzpicture}[ 
    state/.style={
      circle,
      draw=black,
      font=\sf\footnotesize,
      inner sep=1,
      text centered},
    every edge/.append style={font=\sf\footnotesize}
]    
\node[state] (S0) at (0,0.7) {0};
\node (xx) at (-0.8,0.7) [left] {$\sf EX_1$:};
\node[state] (S1) at (1.8,0.7) {1};
\node[state] (S2) at (-0.3,-0.4){2};
\node[state] (S3) at (3.6,0.7) {3};
\node[state] (S4) at (2.7,-0.4) {4};
\node[state] (S5) at (5,0.7) {5};
\node[state] (S6) at (6.4,0.7) {6};

\path[->] (S0) edge node[above] {$(0.5) a$} (S1);
\path[->] (S0) edge node[below right] {$(0.5) a$} (S2);
\path[->] (S1) edge node[above] {$(0.1) b$} (S3);
\path[->, bend left = 30] (S1) edge node[below right] {$(0.9) b$} (S0);
\path[->, bend left = 30] (S2) edge node[left] {$b$} (S0);

\path[->] (S3) edge node[right] {$(0.9) \tau$} (S4);
\path[->] (S3) edge node[above] {$(0.1) \tau$} (S5);
\path[->] (S4) edge node[above] {$a$} (S1);
\path[->] (S5) edge node[above] {$c$} (S6);

\end{tikzpicture}
\end{center}
\vspace{-5mm}
\caption{ An example of a probabilistic model of some SUT called $\sf EX_1$.} \label{fig.probEX1}\label{fig.runningexample}
\end{figure}

\vspace{-3mm}
%
As a running example, consider the labelled transition system (LTS) \cite{arnold1994finite} in
Fig. \ref{fig.runningexample} as a model of some SUT.
The transitions are labelled with actions, e.g. $a$ and $b$. A non-$\tau$
action represents an interaction between the SUT and its environment. In our set up such an action is assumed to
occur {\em synchronously} a la CSP \cite{CSP2004} (for an action $a$ to take place, both the SUT and the 
environment first need to agree on doing $a$; then they will do $a$ together).
The action $\tau$ represents an internal action by the SUT, that is not visible to the environment.

%
%
%
%

%

To test the SUT, the tester controls the SUT by insisting on which action it wants to synchronize; e.g.
if on the state $t$ the SUT is supposed to be able to either do $a$ or $b$, the tester can insist on doing $a$.
If the SUT fails to go along with this, it is an error.
The tester can also test if in this state the SUT can be coerced to do an action that it is not
supposed to synchronize; if so, the SUT is incorrect.
We will assume a black box setup. That is, the tester cannot actually see the SUT's state, though he/she
can try to infer this based on information visible to him/her, e.g. the trace of the external actions
done so far. For example after doing $a$ on the SUT $\sf EX_1$ above,
the tester cannot tell whether it then goes to the state 1 or 2. 
However, if the tester manages to do $abc$ he/she would on the hind sight know
that the state after $a$ must have been 1. 


When a state $s$ has multiple outgoing transitions with the same label, e.g. $a$, this implies non-determinism,
since the environment cannot control which $a$ the SUT will take (the environment can only control
whether or not it wants to do $a$). We assume the modeler is able
estimate the probability of taking each of these $a$-transition and 
annotate this on the transition. 
E.g. in Fig. \ref{fig.runningexample} 
we see that in state 1, two $a$-transitions are possible, leading to different states,
each with the probability of 0.5. 
Similarly, in state 3 there are two $\tau$-transitions leading to states 4 and 5,
with the probability of 0.9 and 0.1 respectively.
A probabilistic model such as in Figure \ref{fig.runningexample}
is also called a Markov Decision Process (MDP) \cite{baier2008principles}. 

Let in the sequel $M$ be an MDP model, with finite number of transitions,
and a single initial state. Let $s,t$ be states, and $a$ an action. 
We write $s {\in} M$ to mean that $s$ is a state in $M$. The notation 
$\transition{s}{a}{t}$ denotes a transition that goes from the state $s$ to $t$ and is labelled with $a$.
We write $\transition{s}{a}{t}\in M$ to mean that $\transition{s}{a}{t}$ is a transition 
in $M$.
$P_M(\transition{s}{a}{t})$ denotes the probability that $M$ will take this particular
transition when it synchronizes over $a$ on the state $s$.

To simplify calculation over non-deterministic actions,
we will assume that $M$ is {\em $\tau$-normalized} in the following sense. 
First, a state cannot have a mix of $\tau$ and non-$\tau$ outgoing transitions.
E.g. a state $s$ with two transitions $\{ \transition{s}{\tau}{t}, \transition{s}{a}{u} \}$
should first be re-modelled as $\{ \transition{s}{\tau}{t}, \transition{s}{\tau}{s'}, \transition{s'}{a}{u} \}$
by introducing an intermediate state $s'$, and the modeler should provide estimation
on the probability of taking each of the two $\tau$ transitions.
\HIDE{
a model should be as such that if on a state $s$ it is possible 
to do a $\tau$ transition, 
then {\em all} outgoing transitions on $s$ should be $\tau$ transitions. For example, a model 
such as $M_0$ below  is excluded from our consideration. Such a model implies that the SUT can 
either internally decide to synchronize on $a$, or to reject it. Model $M_1$ equivalently 
captures this decision point, and should be used instead of $M_0$ (in other words, 
$M_0$ should first be normalized to $M_1$, and the modeler should now specify or estimate what
the probability $p$ should be.).

\begin{center}
\begin{tikzpicture}[ 
    state/.style={
      circle,
      draw=black,
      font=\sf\footnotesize,
      inner sep=1,
      text centered},
    every edge/.append style={font=\sf\footnotesize}
]    
\node[state] (S0a) at (0,0) {};
\node (xxa) at (-0.5,0) {$M_0$:};
\node[state] (S1a) at (1,0.3)   {};
\node[state] (S2a) at (1,-0.3)  {};

\node[state] (S0b) at (3,0) {};
\node (xxb) at (2.5,0) {$M_1$:};
\node[state] (S1b)  at (4.5,0.3) {};
\node[state] (S11b) at (5.3,0.3) {};
\node[state] (S2b) at (4.5,-0.3) {};

\path[->] (S0a) edge node[above] {$a$} (S1a);
\path[->] (S0a) edge node[below] {$\tau$} (S2a);

\path[->] (S0b) edge node[above] {$(p)\; \tau$} (S1b);
\path[->] (S1b) edge node[above] {$a$} (S11b);
\path[->] (S0b) edge node[below] {$(1{-}p)\; \tau$} (S2b);

\end{tikzpicture}
\end{center}
}
Second, $M$ should have no state whose all incoming and outgoing transitions are
$\tau$ transitions. Such a state is considered not interesting for our analyses.
Third, $M$ should not contain a cycle that consists of only $\tau$ transitions.
In a $\tau$-normalized model, non-determinism can only be introduced 
if there is a state $s$ with multiple outgoing
transitions labelled by the same action (which can be $\tau$).
 


We define an {\em execution} of the SUT as a finite path $\rho$ through the model
starting from its initial state. A {\em trace} is a finite sequence of external actions.
The trace of $\rho$, ${\sf tr}(\rho)$, is the sequence external actions induced by $\rho$. 
A {\em legal trace} is a trace that can be produced by some execution of the SUT.
A {\em test-case} is abstractly modeled by a trace. We will restrict to test-cases
that form legal traces,
\HIDE{\footnote{
Here, we exclude negative tests (to see how the SUT responds to 
illegal sequence of actions). 
Note that such test cases can still be considered
in our setup by first introducing an error state in the model, and extending every state with 
illegal actions leading to the error state.}}
e.g. $ab$, $aba$, and $ababc$ are possible test cases for $\sf Ex1$ in Fig. \ref{fig.probEX1}.
A set of test cases is also called a {\em test suite}. 
\HIDE{
In this paper, we will not concern ourself with how the tester produces a test suite.
We assume that we are given a test case or a test suite; our focus is on analyzing their coverage.  
}
Since the model can be non-deterministic, the same test case may trigger multiple possible executions
which are indistinguishable from their trace.
If $\sigma$ is a trace, 
${\sf exec}(\sigma)$ denotes the set of all executions $\rho$ such that ${\sf tr}(\rho) {=} \sigma$, and moreover each such $\rho$ is $\tau$-maximal: it cannot be extended without breaking the property ${\sf tr}(\rho) {=} \sigma$.

\vspace{-4mm}
\subsection{Representing a test case: execution model}

\begin{figure}
\begin{center}
\vspace{-4mm}
\begin{tikzpicture}[ 
    state/.style={
      rectangle,
      rounded corners,
      draw=black,
      font=\sf\footnotesize,
      minimum height=4mm,
      minimum width=7mm,
      inner sep=1,
      text centered},
    every edge/.append style={font=\sf\tiny}
]    
\node[state] (S0x0) at (0,0)    {$u_0 (0)$};
\node[state] (S1x1) at (1.8,0.6)  {$u_1 (1)$};
\node[state] (S1x2) at (1.8,-0.6) {$u_2 (2)$};

\node[state] (S2x0) at (4,-0.6) {$u_4 (0)$};
\node[state] (S2x3) at (3.2,0.6) {$u_3 (3)$};

\node[state] (S3x2) at (6,-0.6) {$u_7 (2)$};
\node[state] (S3x4) at (4.6,0.6)  {$u_5 (4)$};

\node[state] (S4x1) at (6,0.6) {$u_6 (1)$};
\node[state] (End)  at (7.3,0) {$u_8 (\sharp)$};

\path[->] (S0x0) edge node[above] {$(0.5) a$} (S1x1);
\path[->] (S0x0) edge node[below] {$(0.5) a$} (S1x2);
\path[->] (S1x2) edge node[below] {$b$} (S2x0);

\path[->] (S1x1) edge node[above] {$(0.1) b$} (S2x3);
\path[->] (S1x1) edge node[left] {$(0.9) b$} (S2x0);

\path[->] (S2x0) edge node[below] {$(0.5) a$} (S3x2);
\path[->] (S2x0) edge node[below] {$(0.5) a$} (S4x1);

\path[->] (S2x3) edge node[above] {$\tau$} (S3x4);

\path[->] (S3x4) edge node[above] {$a$} (S4x1);

\path[->] (S3x2) edge (End);
\path[->] (S4x1) edge (End);
\end{tikzpicture}
\end{center}
\vspace{-8mm}
\caption{The execution model of the test case $aba$ on $\sf EX_1$.} \label{fig.B.aba}
\end{figure}
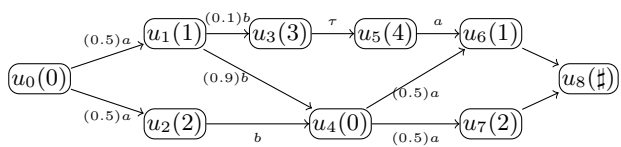

\vspace{-2mm}
The probability that a test case $\sigma$ covers some goal $\phi$ (e.g. a particular state $s$)
can in principle be calculated by quantifying over ${\sf exec}(\sigma)$.
However, if $M$ is highly non-deterministic, the size of ${\sf exec}(\sigma)$
can be exponential with respect to the length of $\sigma$. To facilitate
more efficient coverage calculation we will represent $\sigma$ with the subgraph of $M$
that $\sigma$ induces, called the {\em execution model} of $\sigma$, denoted by ${\sf E}(\sigma)$.
${\sf E}(\sigma)$ forms  a Markov chain;
each branch in ${\sf E}(\sigma)$ is annotated with the probability of taking the branch,
under the premise that $\sigma$ has been observed.
Since a test case is always of finite length and $M$ is assumed to have no $\tau$-cycle,
${\sf E}(\sigma)$ is always acyclic.
Typically the size or ${\sf E}(\sigma)$ (its number of nodes) is much less than the size of ${\sf exec}(\sigma)$.
For example, the execution model of the test case $aba$ on $\sf EX_1$ is shown in Fig. \ref{fig.B.aba}.
An artificial state denoted with $\sharp$ is added so that ${\sf E}(\sigma)$ has a single exit node, 
which is convenient for later.

\vspace{-2mm}
To identify the states in ${\sf E}(\sigma)$ we assign IDs to them
($u_0 ... u_8$ in Fig. \ref{fig.B.aba}). We write
$u.{\sf st}$ to denote $u$'s state label, which is the ID of a state in $M$ that $u$ represents
(so, $u.{\sf st} \in M$); in Fig. \ref{fig.B.aba} this is denoted by the number between brackets in every node.

Importantly, notice that the probability of the transitions in ${\sf E}(\sigma)$ may be different than 
the original probability in $M$. For example, the transition $\transition{u_3}{\tau}{u_5}$
in the above execution model has probability 1.0, whereas in the original model $\sf EX_1$ this corresponds
to the transition $\transition{3}{\tau}{4}$ whose probability is 0.9.
This is because the alternative $\transition{3}{\tau}{5}$ could not have taken place, as it leads
to an execution whose trace does not correspond to the test case $aba$ (which is assumed to have 
happened).

More precisely, when an execution in the model ${\sf E}(\sigma)$ reaches a node $u$,
the probability of extending this execution with the transition $\transition{u}{\alpha}{v}$
can be calculated by taking the conditional probability of the corresponding
transition in the model $M$, given that only the outgoing transitions specified by ${\sf E}(\sigma)$ could happen. 
So,
\HIDE{
\begin{equation}
    P_{{\sf E}(\sigma)}(\transition{u}{\alpha}{v})
       = P_M(\transition{u.{\sf st}}{\alpha}{v.{\sf st}} \; | \; {\sf out}_{{\sf E}(\sigma)}(u))
\end{equation}    
where ${\sf out}_{{\sf E}(\sigma)}(u)$ is the set of $u$'s outgoing transitions in 
${\sf E}(\sigma)$ (they all will have the same label as $\alpha$).
If we work this out, the probability we should assign to every transition $\transition{u}{\alpha}{v}$ in
${\sf E}(\sigma)$ is:
}
\HIDE{   
\begin{equation}
    P_{{\sf E}(\sigma)}(\transition{u}{\alpha}{v})
       = \frac{P_M(\transition{u.{\sf st}}{\alpha}{v.{\sf st}})}{
              \sum_{w \; \mbox{s.t.} \; \transition{u}{\alpha}{w} \in {\sf E}(\sigma)} P_M(\transition{u.{\sf st}}{\alpha}{w.{\sf st}})
              } 
\end{equation}
}
$P_{{\sf E}(\sigma)}(\transition{u}{\alpha}{v})$ is $P_M(\transition{u.{\sf st}}{\alpha}{v.{\sf st}})$
divided by the the sum of $P_M(\transition{u.{\sf st}}{\alpha}{w.{\sf st}})$
of all $w$ such that $\transition{u}{\alpha}{w} \in {\sf E}(\sigma)$.

Let $E = {\sf E}(\sigma)$.
Since $E$ is thus acyclic, the probability that SUT traverses a path/ execution $\rho$ in ${\sf E}(\sigma)$ when it is given $\sigma$
can be obtained by multiplying the probability of all the transitions in the path:
\begin{equation} \label{prog.path}
    P_E(\rho) \ = \ \prod_{\transition{s}{\alpha}{t} \in \rho} P_E(\transition{s}{\alpha}{t})
\end{equation}
%

\vspace{-8mm}
\subsubsection*{Simple coverage analyses.}

As an example of a simple analysis, let's calculate the probability that a test case $\sigma$ produces an execution 
that passes through a given state $s$,
denoted by $P(\WORD{s} \ON \sigma)$. This would then just be the sum of the probability of all full executions 
in ${\sf E}(\sigma)$ that contain $s$. So:

\begin{equation} \label{simple.query.formula}
    P(\WORD{s} \ON \sigma) \ = \ \sum_{\rho \; \mbox{s.t.}\; \rho\in {\sf E}(\sigma) \wedge s\in \rho} P_{{\sf E}(\sigma)}(\rho) 
\end{equation}

For example, on the execution model $\sf EX_1$, $P(\WORD{1} \ON aba) = 0.75$,
$P(\WORD{2} \ON aba) = 0.725$, $P(\WORD{4} \ON aba) = 0.05$,
whereas $P(\WORD{5} \ON aba) = 0$.

\HIDE{
\[
\scriptsize
\begin{array}{cc}
\begin{array}[t]{|c|c|} \hline
\mbox{coverage goal} & \mbox{probability $P(goal \ON aba)$} \\ \hline
\WORD{0} & 1.0 \\ \hline
\WORD{1} & 0.75 \\ \hline
\WORD{2} & 0.725 \\ \hline
\end{array}
&
\begin{array}[t]{|c|c|} \hline
\mbox{coverage goal} & \mbox{probability $P(goal \ON aba)$} \\ \hline
\WORD{3} & 0.05 \\ \hline  
\WORD{4} & 0.05 \\ \hline
\WORD{5} & 0.0 \\ \hline
\end{array}
\end{array}
\]
}

\section{Coverage under Uncertainty} \label{sec.cov}

\vspace{-3mm}
Coverage goals posed in practice are however more complex than goals
exemplified above. Let us first introduce a language for expressing 'goals'; 
we will keep it simple, but expressive enough to express what is latter called
'aggregate $k$-wise' goals.
A goal of the form $\WORD{0,2,0}$ is called a {\em word}, 
expressing an intent to cover the subpath $\WORD{0,2,0}$ in the MDP model.
We will also allow disjunctions of words
and sequences of words (called {\em sentences}) to appear as goals. 
For example:
%
$(\WORD{0,2} \vee \WORD{1,0}) \; ; \; \WORD{1}$
formulates a goal to first cover the edge $\shorttransition{0}{2}$ or $\shorttransition{1}{0}$,
and then (not necessarily immediately) the node 1. 

The typical goal people have in practice is to cover at least $p\%$ of the states.
This is called an {\em aggregate goal}. We write this a bit differently: 
a goal of the form $^1{\geq}N$ express an intent to cover at least $N$ different states. 
Covering at least $p\%$ can be expressed as $^1{\geq}\lfloor p*K/100 \rfloor$ where
$K$ is the number of states in the model.
To calculate probabilistic coverage in $k$-wise  testing \cite{ammann2016introduction}, 
the goal $^k{\geq}N$ expresses an intent to cover at least $N$ different words of length $k$.
Now, more formally:

\HIDE{
When considering the coverage of a test suite, we may want to inspect more complicated
coverage goals. For example, if we apply pair-wise testing \cite{ammann2016introduction}, we may want to know if the test covers certain
pairs of adjacent states (rather than just a single state as in the previous example),
or more generally if it covers certain segments of connected states of length $k$.
We may want to know if the test manages to cover certain segments in a particular order.
Finally, we may want to know if the test manage to cover all possible segments
of length $k$ (or at least to know the probability of this).
}


\begin{mydef}
A coverage goal is a formula $\phi$ with this syntax:    
\[ \begin{array}{lclr}
    \phi & ::= & S \ | \ A  & \mbox{(goal)}\\
    S & ::= & C \ | \ C ; S & \mbox{(sentence)} \\
    A & ::= & ^k{\geq} N    & \mbox{(aggregate goal), with $k{\geq}1$} \\
    C & ::= & W \ | \ W {\vee} C  & \mbox{(clause)}\\
    W & ::= & \WORD{s_0,...,s_{k{-}1}} & \mbox{(word), with $k{\geq}1$} 
   \end{array}
\]
\end{mydef}

A sentence is a sequence $C_0 ; C_1 ;  ...$. Each $C_i$ is called a {\em clause}, 
which in turns consists of one or more words.
A {\em word} is denoted by $\WORD{s_0,s_1,...}$ and specifies one or more 
connected states in an MDP. 

Let $\rho$ be an execution. If $\phi$ is a goal, we write 
$\rho \vdash \phi$ to mean that $\rho$ covers $\phi$.
Checking this is decidable.
For a word $W$, $\rho \vdash W$ if $W$ is a segment of $\rho$.
For a clause $C = W_0 \vee W_2 \vee ...$, $\rho \vdash C$ if $\rho \vdash W_k$
for some $k$.
Roughly, a sentence $C_0 ; C_1 ; ...$ is covered by $\rho$ if all clauses $C_i$
are covered by $\rho$, and furthermore they are covered in the order
as specified by the sentence. We will however define it more loosely to allow consecutive
clauses to overlap, as follows:

\begin{mydef}[Sentence Coverage] \label{def.sentence.cov}
Let $S$ be a sentence. (1) An empty $\rho$ does not cover $S$.
(2) If $S$ is a just a single clause $C$, then $\rho \vdash S$ iff $\rho \vdash C$.
(3) If $S = C ; S'$ and a prefix of $\rho$ matches one of the words in $C$,
  then $\rho \vdash S$ iff $\rho \vdash S'$.
  If $\rho$ has no such prefix, then $\rho \vdash S$ iff ${\sf tail}(\rho) \vdash S$.
\end{mydef}

An aggregate goal of the form $^k{\geq} N$ is covered by $\rho$ if $\rho$
covers at least $N$ {\em different} words of size $k$. While sentences are
expressible in temporal logic, aggregate goals are not.
This has an important consequence discussed later.



\HIDE{
Sentences are expressible in CTL.
Therefore, their probabilistic coverage can be calculated
through probabilistic CTL (PCTL) model checking \cite{hansson1994logic,kwiatkowska2007stochastic,baier2008principles}.

A full until-operator is not used in our goal language. 
Also, unlike in CTL, we do {\em not} allow
future operators to recurse on their left operand; e.g. $(C_1 ; C_2) ; C_3$ is not
allowed. 
We do not think it to be of much use as a coverage goal, and excluding it
simplifies the algorithm in Section \ref{sec.efficient.cov.calc} 
for calculating probabilistic coverage.

However, {\em aggregate goals have no counterpart in CTL nor LTL}.
}


Let $\phi$ be a coverage goal and $\sigma$ a test case.
Let's write $P(\phi \ON \sigma)$ to denote the probability
that $\phi$ is covered by $\sigma$, which 
can be calculated analogous to (\ref{simple.query.formula}) as follows: 

\begin{mydef} \label{def.tc.query.prob}
$P(\phi \ON \sigma)$ is equal to $P(\phi \ON E)$ where $E = {\sf E}(\sigma)$,
$ 
P(\phi \ON E)\ = \ \sum_{\rho \; \mbox{s.t.} \; \rho\in {\sf exec}(E) \; \wedge \; \rho \vdash \phi} P_E(\rho)
$,
and where $P_E(\rho)$ is calculated as in (\ref{prog.path}). 
\end{mydef}

For example, consider the test case $aba$ on the SUT ${\sf EX}_1$. Fig. \ref{fig.B.aba} shows the
execution model of $aba$. 
$P(\WORD{2,0} \ON aba)$ is the probability that $aba$'s execution passes
through the transition $\shorttransition{2}{0}$; this probability is 0.5.
$P((\WORD{2} \vee \WORD{3}) ; \WORD{1} \ON aba)$ is the probability that 
$aba$ first visits the state 2 or 3, and sometime later 1; this probability is 0.75.
$P(^{1}{\geq}4 \ON aba)$ is the probability
that the execution of $aba$ visits at least four different states; this is unfortunately only 0.05.

Due to non-determinism, the size of ${\sf exec}(\sigma)$ could be exponential
with respect to the length of $\sigma$. 
Simply using
using the formula in Def. \ref{def.tc.query.prob} would then be expensive.
Below we present a much better algorithm to do the calculation.

\section{Efficient coverage calculation} \label{sec.efficient.cov.calc}


\vspace{-3mm}
Coverage goals in the form of sentences are actually expressible in Computation Tree Logic (CTL)
\cite{baier2008principles}. E.g. $\WORD{s,t};\WORD{u}$ corresponds to
${\sf EF}(s\wedge t \wedge {\sf EF} u)$. It follows, that the probability of
covering a sentence can be calculated through 
probabilistic CTL model checking \cite{hansson1994logic,baier2008principles}.
\HIDE{
(strictly speaking, the latter does not calculate the probability of a property $\phi$,
but rather whether or not $\phi$ has the probability of e.g. at least some $p$.
However, under the hood does calculates the probability of $\phi$, which
is the information that we need here.). 
}
Unfortunately, aggregate goals are not expressible CTL. Latter we will discuss 
a modification of probabilistic model checking to allow the calculation of
aggregate goals. We first start with the calculation of {\em simple sentences}
whose words are all of the length one.

Let $S$ be a simple sentence, $\sigma$ a test case, and $E={\sf E}(\sigma)$. 
In standard probabilistic model checking, $P(S | \sigma)$ would be calculated through
a series multiplications over a probability matrix \cite{baier2008principles}. 
We will instead do it by performing labelling on the nodes of 
$E$, resembling more to non-probabilistic CTL model checking. This approach
is more generalizable to later handle aggregate goals. 

Notice that any node $u$ in $E$ induces a unique
subgraph, denoted by $E@u$, rooted in $u$. It represents the remaining execution of $\sigma$, starting
at $u$. 
When we label $E$ with some coverage goal
$\psi$, the labelling will proceed in such a way that when it terminates every node $u$
in $E$ is extended with labels of the form $u.{\sf lab}(\psi)$ containing the value of 
$P(\psi \ON E@u)$.
The labelling algorithm is shown in Fig. \ref{fig.labeling.alg}, namely the procedure ${\sf label}(..)$ ---we will explain it below.
In any case, after calling ${\sf label}(E,S)$, 
the value of $P(S \ON \sigma)$ can thus be obtained simply by inspecting the ${\sf lab}(S)$
of $E$'s root node. This is done by the procedure $\sf calcSimple$.

\begin{figure}
\vspace{-5mm}
\begin{center}
\begin{multicols}{2}  
\begin{algorithmic}[1]
    \Procedure{\sf calcSimple}{$E,S$}
       \State ${\sf label}(E,S)$
       \State {\bf return} \; ${\sf root}(E).{\sf lab}(S)$
    \EndProcedure
    \Statex
    \Procedure{\sf label}{$E,S$}
        \State $u_0 \gets {\sf root}(E)$
        \State ${\bf case} \; S \; {\bf of}$
            \State \ \ $C \hspace{6mm} \rightarrow {\sf label1}(u_0,C)$
            \State \ \ $C ; S' \ \rightarrow
                          {\sf label}(E,S') \; ; \; 
                          {\sf label1}(u_0, S) $
    \EndProcedure
    \Statex
    \Procedure{\sf checkClause}{$u,C$}
       \State {\footnotesize\it  $\rhd$ the clause $C$ is assumed to be of this form, with $k\geq 1$} : 
       \State {\bf let}\; $\WORD{s_0} \vee ... \vee \WORD{s_{k{-}1}} = C$
       \State $isCovered \gets u.{\sf st} \in \{s_0, ... , s_{k{-}1} \}$
       \State {\bf return}\; $isCovered$
    \EndProcedure
    \Statex
    \Procedure{\sf label1}{$u,S$}
    \State {\footnotesize \it $\rhd$ recurse to $u$'s successors} :
    \State ${\bf forall}\; v \in u.{\sf next} \ \rightarrow \ {\sf label1}(v,S)$
    \State {\footnotesize  \it $\rhd$ pre-calculate $u$'s successors' total probability to cover $S$} : 
    \State $q' \gets  \sum_{v {\in} u.{\sf next}} u.{\sf pr}(v)*v.{\sf lab}(S)$
    \State {\footnotesize\it $\rhd$ calc. $u$'s probability to cover $S$} : 
    \State ${\bf case} \; S \; {\bf of}$
      \State \ \ $C \hspace{6mm} \rightarrow 
                      \begin{array}[t]{l}
                      {\bf if}\; {\sf checkClause}(u,C)  \\
                      {\bf then} \; q \gets 1 \\
                      {\bf else} \; q \leftarrow q'
                      \end{array} $
    
       \State \ \ $C ; S' \ \rightarrow
                      \begin{array}[t]{l}
                      {\bf if}\; {\sf checkClause}(u,C) \\
                      {\bf then} \; q \gets u.{\sf lab}(S') \\
                      {\bf else} \; q \gets q' 
                      \end{array} $
    
    \State ${\bf end\;case}$
    \State {\footnotesize\it  $\rhd$ add the calculated probability as a new label to $u$} : 
    \State $u.{\sf lab}(S) \gets q$
    \EndProcedure
\end{algorithmic} 
\end{multicols}
\end{center}

\vspace{-7mm}
\caption{The labeling algorithm of to calculate the probability of simple sentences.} \label{fig.labeling.alg}
\end{figure}  

\vspace{-5mm}
Since $S$ is a sentence, it is a sequence of clauses.
The procedure ${\sf label}(E,S)$ first recursively labels $E$ with with the tail $S'$ of $S$ (line 9),
then we proceed with the labelling with $S$ itself, which is done by the procedure $\sf label1$.
In $\sf label1$, the following notations are used.
Let $u$ be a node in $E$. Recall that $u.{\sf st}$ denotes the ID
of the state in $M$ that $u$ represents. We write $u.{\sf next}$ to denote the set of 
$u$'s successors in $E$ (and not in $M$!). For such successor $v$,
$u.{\sf pr}(v)$ denotes the probability annotation that $E$ puts on the arrow $\shorttransition{u}{v}$. 
A label is a pair $(\psi,p)$ where $\psi$ is a coverage goal and $p$ is
probability in $[0..1]$. The notation $u.{\sf lab}$ denotes the labels 
put so far to the node $u$. The assignment $u.{\sf lab}(\psi) \leftarrow p$
adds the label $(\psi,p)$  to $u$, and the expression $u.{\sf lab}(\psi)$
returns now the value of $p$. 

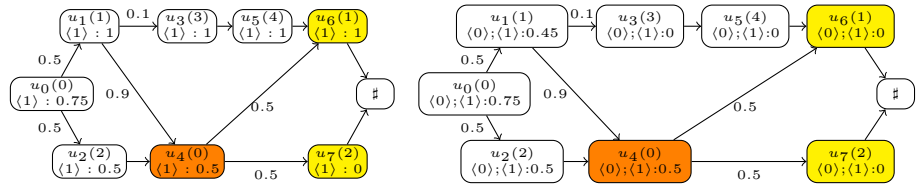
\begin{figure}
\vspace{-7mm}
\begin{center}
\begin{tabular}{cc}
\begin{tikzpicture}[ 
    state/.style={
      rectangle,
      rounded corners,
      draw=black,
      font=\sf\tiny,
      inner sep=0.3,
      text centered},
    every edge/.append style={font=\sf\tiny}
]    
    \node[state] (S0x0) at (0,0)    {$\begin{array}{c} u_0 (0) \\ \WORD{1} : 0.75 \end{array}$};
    \node[state] (S1x1) at (0.5,0.9)  {$\begin{array}{c} u_1 (1) \\ \WORD{1} : 1 \end{array}$};
    \node[state] (S1x2) at (0.5,-0.9) {$\begin{array}{c} u_2 (2) \\ \WORD{1} : 0.5 \end{array}$};

    \node[state] (S2x3)  at (1.8,0.9) {$\begin{array}{c} u_3 (3) \\ \WORD{1} : 1 \end{array}$};
    \node[state, style={fill=orange}] (S2x0) at  (1.8,-0.9) {$\begin{array}{c} u_4 (0) \\ \WORD{1} : 0.5 \end{array}$};

    \node[state] (S3x4) at (2.8,0.9) {$\begin{array}{c} u_5 (4) \\ \WORD{1} : 1 \end{array}$};
    \node[state, style={fill=yellow}] (S4x1) at (3.8,0.9) {$\begin{array}{c} u_6 (1) \\ \WORD{1} : 1 \end{array}$};
    \node[state, style={fill=yellow}] (S3x2) at (3.8,-0.9) {$\begin{array}{c} u_7 (2) \\ \WORD{1} : 0 \end{array}$};

    \node[state, style={minimum height=4mm, minimum width=5mm}] (End) at (4.3,0) {$\sharp$};

    \path[->] (S0x0) edge node[left] {$0.5$} (S1x1);
    \path[->] (S0x0) edge node[left] {$0.5$} (S1x2);
    \path[->] (S1x2) edge node[below] {} (S2x0);

    \path[->] (S1x1) edge node[above] {$0.1$} (S2x3);
    \path[->] (S1x1) edge node[left] {$0.9$} (S2x0);

    \path[->] (S2x0) edge node[below] {$0.5$} (S3x2);
    \path[->] (S2x0) edge node[below] {$0.5$} (S4x1);

    \path[->] (S2x3) edge node[above] {} (S3x4);

    \path[->] (S3x4) edge node[above] {} (S4x1);

    \path[->] (S3x2) edge (End);
    \path[->] (S4x1) edge (End);
    \end{tikzpicture}
\ \ &
    \begin{tikzpicture}[ 
        state/.style={
          rectangle,
          rounded corners,
          draw=black,
          font=\sf\tiny,
          inner sep=2,
          text centered},
        every edge/.append style={font=\sf\tiny}
    ]    
        \node[state] (S0x0) at (0,0)    
           {$\begin{array}{c} u_0 (0) \\ \WORD{0} {;} \WORD{1} {:} 0.75 \end{array}$};
        \node[state] (S1x1) at (0.5,0.9)  
           {$\begin{array}{c} u_1 (1) \\ \WORD{0} {;} \WORD{1} {:} 0.45 \end{array}$};
        \node[state] (S1x2) at (0.5,-0.9) 
           {$\begin{array}{c} u_2 (2) \\ \WORD{0} {;} \WORD{1} {:} 0.5 \end{array}$};

        \node[state] (S2x3)  at (2.2, 0.9) 
             {$\begin{array}{c} u_3 (3) \\ \WORD{0} {;} \WORD{1} {:} 0 \end{array}$};
        \node[state, style={fill=orange}] (S2x0) at  (2.2,-0.9) 
             {$\begin{array}{c} u_4 (0) \\ \WORD{0} {;} \WORD{1} {:} 0.5 \end{array}$};

        \node[state] (S3x4) at (3.6,0.9) 
             {$\begin{array}{c} u_5 (4) \\ \WORD{0} {;} \WORD{1} {:} 0 \end{array}$};
        \node[state, style={fill=yellow}] (S4x1) at (5,0.9) 
             {$\begin{array}{c} u_6 (1) \\ \WORD{0} {;} \WORD{1} {:} 0 \end{array}$};
        \node[state, style={fill=yellow}] (S3x2) at (5,-0.9) 
             {$\begin{array}{c} u_7 (2) \\ \WORD{0} {;} \WORD{1} {:} 0 \end{array}$};

        \node[state, style={minimum height=4mm, minimum width=5mm}] (End) at (5.6,0) {$\sharp$};

        \path[->] (S0x0) edge node[left] {$0.5$} (S1x1);
        \path[->] (S0x0) edge node[left] {$0.5$} (S1x2);
        \path[->] (S1x2) edge node[below] {} (S2x0);

        \path[->] (S1x1) edge node[above] {$0.1$} (S2x3);
        \path[->] (S1x1) edge node[left] {$0.9$} (S2x0);

        \path[->] (S2x0) edge node[below] {$0.5$} (S3x2);
        \path[->] (S2x0) edge node[below] {$0.5$} (S4x1);

        \path[->] (S2x3) edge node[above] {} (S3x4);

        \path[->] (S3x4) edge node[above] {} (S4x1);

        \path[->] (S3x2) edge (End);
        \path[->] (S4x1) edge (End);
        \end{tikzpicture}
    \end{tabular}
    \end{center}
    
    \vspace{-5mm}
    \caption{The {\bf left graph} shows the result of ${\sf label}(\WORD{1})$ on the execution model
    of $aba$ in Fig.\ref{fig.B.aba}. For simplicity, the action labels on the arrows are removed.
    The probability annotation is kept.
    In turn, ${\sf label}()$ calls $\sf label1$, which then
    performs the labelling recursively from right to left. The nodes $u_6$ and
    $u_7$ (yellow) are base cases. The probabilities of $\WORD{1}$ on them are respectively
    1 and 0. This information is then added as the labels of these nodes.
    Next, $\sf label1$ proceeds with the labelling of $u_4$ and $u_5$. E.g. on $u_4$
    (orange), because $u_4.{\sf st}$ is not 1, for $u_4$ to cover $\WORD{1}$ we need an execution
    that goes through $u_6$, with the probability of 0.5. So the probability of $\WORD{1}$ on $u_4$
    is 0.5.
    The {\bf right graph} shows the result of ${\sf label}(\WORD{0} {;} \WORD{1})$ on the same
    execution model. This will first call ${\sf label}(\WORD{1})$, thus producing the labels as shown
    in the left graph, then proceeds with ${\sf label1}(\WORD{0} {;} \WORD{1})$. Again,
    $\sf label1$ performs the labelling recursively from right to left. The base cases
    $u_6$ and $u_7$ do not cover $\WORD{0} {;} \WORD{1}$, so the corresponding
    probability there is 0. Again, this information is added as labels of the corresponding nodes.
    Node $u_4$ (orange) has $u_4.{\sf st} = 0$. So, any execution that starts from there and covers
    $\WORD{1}$ would also cover $\WORD{0} {;} \WORD{1}$. The probability that $u_4$
    covers $\WORD{1}$ is already calculated in the left graph, namely 0.5. So this is also the
    probability that it covers $\WORD{0} {;} \WORD{1}$.
    } \label{fig.example.labeling}
\end{figure}

The procedure ${\sf label1}(\psi)$ will perform the labelling node by node
recursively in the bottom-up direction over the structure of $E$ (line 19).
Since $E$ is acyclic, only a single pass of this recursion is needed.
For every node $u \in E$, ${\sf label1}(u,S)$ has to add a new
label $(S,q)$ to the node $u$ where $q$ is the probability
that the goal $S$ is covered by the part of executions of $\sigma$ that starts
in $u$ (in other words, the value of $P(S \ON E@u)$). 
The goal $S$ will be in one of these two forms:

\begin{enumerate}
    \item $S$ is just a single clause $C$ (line 24). Because $S$ is a simple sentence, $C$ is
    a disjunction of singleton words
    $\WORD{s_0} \vee ... \vee \WORD{s_{k{-}1}}$, where each $s_i$ is an ID of a state in $M$. 
    If $u$ represents one of these states, the probability that
    $E@u$ covers $C$ would be 1. Else, it is the sum of the probability
    to cover $C$ through $u$'s successors (line 20).
    As an example, Fig. \ref{fig.example.labeling} (left) shows how the labeling of a simple sentence
    $\WORD{1}$
    on the execution model in Fig. \ref{fig.B.aba} proceeds.
     
    \item $S$ is a sentence with more than one clauses; so it is of the form $C;S'$ (line 25) where $C$ is a clause
    %
    and $S'$ is the rest of the sentence, we calculate the coverage probability by $E@u$ by
    basically following the third case in Def. \ref{def.sentence.cov}.
    
    As an example, Fig. \ref{fig.example.labeling} (right) shows how the labeling of $S = \WORD{0} {;} \WORD{1}$
    proceeds. At every node $u$ we first check if $u$ covers the first word, namely $\WORD{0}$.
    If this is the case, the probability that $E@u$ covers $S$ would be
    the same as the probability that it covers the rest of $S$, namely $\WORD{1}$.
    The probability of the later is by now known, calculated by $\sf label$
    in its previous recursive call. The result can be inspected in $u.{\sf lab}(\WORD{1})$.
    
    If $u$ does {\em not} cover $S$, the probability that ${\sf E}(u)$ covers $S$ 
    would be the sum of the probability to cover $S$ through $u$'s successors 
    (calculated in line 21).
        
\end{enumerate}

Assuming that checking if a node locally covers a clause (the procedure $\sf checkClause$ in Fig. \ref{fig.labeling.alg})
takes a unit time, the time complexity of $\sf label1$ is 
${\mathcal O}(|E|)$, where $|E|$ is the size of $E$ in terms of its the number
of edges. The complexity of $\sf label$ is thus ${\mathcal O}(|E|*|S|)$, where $|S|$ is the size of the goal $S$ in terms of the number of clauses it has. The size of $E$ is typically just linear to the length of the test case: ${\mathcal O}(N_{sucs}*|\sigma|)$,
where $N_{sucs}$ is the average number of successors that each state in $M$ has.
This is significant improvement compared to the exponential
run time that we would get if we simply use Def. \ref{def.tc.query.prob}.
%
%

\vspace{-5mm} 
\subsection{Non-simple sentences}  \label{sec.word.expansion}
 
\vspace{-3mm} 
Coverage goals in $k$-wise testing would require sentences with words of length $k{>}1$ to express.
These are thus {\em non-simple} sentences. We will show that the algorithm in Fig. \ref{fig.labeling.alg}
can be used to handle these sentences as well.

\HIDE{
To handle all types of non-aggregate goals we can opt to generalize the algorithm in Fig. \ref{fig.labeling.alg}
so that it also works on non-simple sentences. In particular,
${\sf checkClause}(u,C)$ needs to be modified. If the words in $C$ are all of length 1,
then either $C$ will hold on $u$ or it does not. If some words have length ${>}1$,
then $C$ may hold on $u$ with some probability. The probability of a word $w$ of length $k{>}1$ to hold on $u$
can be calculated by quantifying over all outgoing paths originating from $u$ and of length $k$ to check if
$w$ is one of these paths, and if so we calculate its probability using (\ref{prog.path}).
We have to do this for every word in $C$, and then sum the probabilities to obtain the probability of $C$
on $u$. 
We will however do this differently, with as the benefit that there is no need to change the algorithm in Fig. \ref{fig.labeling.alg}.
}

\HIDE{
Note this approach means that for every word that occur in our coverage goal, we will have
to quantify over all outgoing paths of every node, up to a certain length $k$. Alternatively,
which is what we are going to do, we can pre-calculate these outgoing paths, because then we only
to construct them once.
}

Consider as an example the sentence $\WORD{0,2,0} ; \WORD{4,1,\sharp}$. The words are of length three,
so the sentence is non-simple.
Suppose we can treat these words as if they are singletons. E.g. in $\WORD{0,2,0}$ 
the sequence $0,2,0$ is treated as a single symbol, and hence the word is a singleton. 
From this perspective, any non-aggregate goal is thus a simple sentence, and 
therefore the algorithm in  Fig. \ref{fig.labeling.alg} can be used to calculate its
coverage probability. We do however need to pre-process the execution model to align it
with this idea.

The only part of the algorithm in Fig. \ref{fig.labeling.alg} where the size of the words matters 
is in the procedure ${\sf checkClause}$. Given a node $u$ in the given execution model $E$  and a clause
$C$, ${\sf checkClause}(u,C)$ checks if the clause $C$ is covered by $E$'s executions that
start at $u$. If the words in $C$ are all of length one, $C$ can be immediately checked
by knowing which state in $M$ $u$ represents. This information is available in the attribute
$u.{\sf st}$. Clauses with longer words can be checked in a similar way. For simplicity,
assume that the words are all of length $k$. We first restructure
$E$ such that the ${\sf st}$ attribute of every node $u$ in the new $E$ contains a word of length $k$
that would be covered if the execution of $E$ arrives at $u$. We call this restructuring
step {\em $k$-word expansion}. Given a base execution model $E$, the produced
new execution model will be denoted by $E^k$. 
As an example, the figure below shows the word expansion with $k{=}3$ of the execution model in Fig. \ref{fig.B.aba} (for every node $v$
we only show its $v.{\sf st}$ label, which is an execution segment of length 3).
Artificial initial and terminal states are added to the new execution model, labelled with $\sharp$.
When a word of length $k$ cannot be formed, because the corresponding segment has reached the terminal
state $\sharp$ in $E$, we pad the word with $\sharp$'s on its the end until
its length is $k$. 

\begin{center}
\vspace{-3mm}
\begin{tikzpicture}[ 
    state/.style={
      rectangle,
      rounded corners,
      draw=black,
      font=\sf\tiny,
      text centered},
    every edge/.append style={font=\sf\tiny}
]    
\node[state, style={fill=orange,minimum width=5mm}] (Start) at (0,0)     {$\sharp$};

\node[state, style={fill=yellow}] (S013) at (1.5,1)  {[0,1,3]};
\node[state] (S134) at (3,1)  {[1,3,4]};
\node[state] (S341) at (4.5,1)  {[3,4,1]};

\node[state, style={fill=yellow}] (S010) at (1.5,0) {[0,1,0]};
\node[state] (S101)     at (3,0.5)   {[1,0,1]};
\node[state] (S01sharp) at (4.5,0.5) {$[0,1,\sharp]$};
\node[state] (S102)     at (3,0)     {[1,0,2]};
\node[state] (S02sharp) at (4.5,0)   {$[0,2,\sharp]$};

\node[state, style={fill=yellow}] (S020) at (1.5,-1)  {[0,2,0]};
\node[state] (S201)     at (3,-0.5)   {[2,0,1]};
\node[state] (S01sharpX) at (4.5,-0.5) {$[0,1,\sharp]$};
\node[state] (S202)     at (3,-1)     {[2,0,2]};
\node[state] (S02sharpX) at (4.5,-1)   {$[0,2,\sharp]$};
    
\node[state, style={fill=yellow}] (S41) at (6,1)      {$[4,1,\sharp]$};
\node[state, style={fill=yellow}] (S1)  at (6,0)    {$[1,\sharp,\sharp]$};
\node[state, style={fill=yellow}] (S2)  at (6,-1) {$[2,\sharp,\sharp]$};

\node[state, style={fill=orange, minimum width=5mm}] (End) at (7.5,0) {$\sharp$};

\path[->] (Start) edge node[left] {0.05} (S013);
\path[->] (Start) edge node[below] {0.45} (S010);
\path[->] (Start) edge node[left] {0.5} (S020);

\path[->] (S013) edge (S134);
\path[->] (S134) edge (S341);
\path[->] (S341) edge (S41);
\path[->] (S010) edge node[above] {0.5} (S101);
\path[->] (S010) edge node[below] {0.5} (S102);
\path[->] (S101) edge (S01sharp);
\path[->] (S01sharp) edge (S1);
\path[->] (S102) edge (S02sharp);
\path[->] (S02sharp) edge (S2);

\path[->] (S020) edge node[above] {0.5} (S201);
\path[->] (S020) edge node[below] {0.5} (S202);
\path[->] (S201) edge (S01sharpX);
\path[->] (S01sharpX) edge (S1);
\path[->] (S202) edge (S02sharpX);
\path[->] (S02sharpX) edge (S2);

\path[->] (S41) edge (End);
\path[->] (S1)  edge (End);
\path[->] (S2)  edge (End);

\end{tikzpicture}
\end{center}

\vspace{-7mm}
\subsection{Coverage of aggregate goals} \label{subsec.aggregate}

\vspace{-2mm}
We will only discuss the calculation of aggregate goals of the form $^k{\geq}N$
where $k{=}1$. If $k{>}1$ we can first apply a $k$-word expansion (Section \ref{sec.word.expansion})
on the given execution model $E$,  then we calculate $^1{\geq}N$ on the expanded 
execution model.

Efficiently calculating $^{1}{\geq} N$ is more challenging. The algorithm below
proceeds along the same idea as how we handled simple sentences, namely
by recursing over $E$. We first need to extend every node $u$ in
$E$ with a new label $u.{\sf A}$. This label is a set containing pairs
of the form $V \mapsto p$ where $V$ is a set of $M$'s states
and $p$ is the probability that $E@u$ would cover {\em all} the states mentioned in $V$.
Only $V$'s whose probability is non-zero need to be included in this mapping.  
After all nodes in $E$ is labelled like this, the probability $^1{\geq}N$
can be calculated from the ${\sf A}$ of the root node $u_0$:

\begin{equation}
   P(^1{\geq}N \ON \sigma) \ = \
      \sum_{V{\mapsto}p \;\in \; u_0.A} {\bf if}\; |V| \geq N \; {\bf then} \; p \; {\bf else} \; 0 
\end{equation}
     
The labelling is done recursively over $E$ as follows:      
      
\begin{enumerate}
    \item The base case is the terminal node $\#$. The $\sf A$ label
    of $\#$ is just $\emptyset$.
    
    \item For every node $u \in E$, we first recurse to all its successors. Then,
    we calculate a preliminary mapping for $u$ in the following
    {\em multi-set} $A'$:
    \[ 
       A' \ = \
       \{ \; V {\cup} \{ u.{\sf st} \} \mapsto p{*}P_E(\shorttransition{u}{v}) 
          \ | \
          v \in u.{\sf next}, \
          V{\mapsto} p \in v.{\sf A} \; \}
    \] 
    As a multi-set note that $A'$ may contain duplicates, e.g. two instances of $V\mapsto p_0$. Additionally,
    it may contain different maps that belong to the same $V$, e.g. $V\mapsto p_1$ and $V\mapsto p_2$.
    All these instances of $V$ need to be merged by summing up their $p$'s, e.g.
    the above instances is to be merged to $V \mapsto p_0 {+} p_0 {+} p_1 {+} p_2$
    The function $\sf merge$ will do this.
    The label $u.{\sf A}$ is then just:
    $u.{\sf A} \ = \ {\sf merge}(A')$
    
\end{enumerate}   
The recursion terminates because $E$ is acyclic. 

The above algorithm can however perform worse than a direct calculation via Def. \ref{def.tc.query.prob}.
The reason is that $\sf merge$ is an expensive operation if we do it literally at every node. If we do not merge at all,
and make the ${\sf A}$'s multi-sets instead of sets, we will end up with $u_0.{\sf A}$ that contains as
many elements as the number of paths in $E$, so we are not better of either. Effort to merge
is well spent if it delivers large reduction in the size of the resulting set, otherwise the effort is wasted.
Unfortunately it is hard to predict the amount of reduction we would get for each particular merge.
We use the following merge policy.
We only merge on nodes at the $B{-}1$-th position of 'bridges' where $B$ is the length of the bridge at hand.
A bridge is a sequence of nodes $v_0,...,v_{B{-}1}$ such that: (1) every $v_i$ except the last one has only
one outgoing edge, leading to $v_{i{+}1}$, and (2) the last node $v_{B{-}1}$ should have more
than one successor.
A bridge forms thus a deterministic section of $E$, that leads to a non-deterministic
section. Merging on a bridge is more likely to be cost effective. Furthermore, only one merge is needed for an entire bridge. 
Merging on a non-deterministic node (a node with multiple successors) is risky. This policy takes
a conservative approach by not merging at all on such nodes.
The next section will discuss the performance of our algorithm.

\vspace{-3mm}    
\section{Experimental Results} \label{sec.experiment}

\vspace{-3mm}
In the following experiment we benchmark the algorithm from Section \ref{sec.efficient.cov.calc}
against the 'brute force' way to calculate coverage using Def. \ref{def.tc.query.prob}.
We will use a family of models $M_m$ in Fig. \ref{fig.Mk}. Despite its simplicity, 
$M_m$ is highly non-deterministic and is designed to generate a large number
of executions and words. 

We generate a family of execution models $E(i,m)$ by applying a test case $tc^i$ on the model
$M_m$ where $m \in \{0,2,8\}$. The test case is:

\[ tc^i \ = \ a c^i a b^i a c^i a \]

The table in Fig. \ref{fig.experiment.emodels} (left) 
shows the statistics of all execution models used in this experiment.
Additionally we also construct $E(i,m)^3$ (applying 3-word expansion). 
The last column in the table shows the number of nodes in the corresponding
$E(i,m)^3$ (the number of executions stays the same, of course).

\begin{figure}
%
\begin{center}
\begin{tikzpicture}[ 
    state/.style={
      circle,
      draw=black,
      font=\sf\footnotesize,
      inner sep=1,
      text centered},
    every edge/.append style={font=\sf\footnotesize}
]    
\node[state] (S0) at (0,0) {0};
\node (xx) at (-0.4,0) [left] {$M_m:$};
\node[state] (S1) at (1.5,0)  {1};
\node[state] (S2) at (1.5,1){2};
\node[state] (S3) at (3,0) {3};
\node[state] (S4) at (5,0) {4};
\node[state] (T0) at (5,1) {$t_0$};
\node        (Ti) at (5,1.5) {{\Large ...}};
\node[state] (Tk) at (5,2.3) {$t_{m{-}1}$};
\node[state] (DONE) at (7,0) {5};

\path[->] (S0) edge node[below] {$a$} (S1);
\path[->, bend right = 30] (S1) edge node[right] {$(0.3) c$} (S2);
\path (S1) edge [loop below] node[below] {$(0.7) c$} (S1);
\path[->] (S1) edge node[below] {$a$} (S3);

\path[->, bend right = 30] (S2) edge node[left] {$(0.7) c$} (S1);
\path (S2) edge [loop above] node[above] {$(0.3) c$} (S2);

\path (S3) edge [loop below] node[below] {$b$} (S3);
\path[->] (S3) edge node[below] {$a$} (S4);

\path (S4) edge [loop below] node[below] {$(q) c$} (S4);
\path[->, bend right = 30] (S4) edge node[right] {$(p) c$} (T0);
\path[->, bend right = 60] (S4) edge node[right] {$(p) c$} (Tk);
\path[->] (S4) edge node[below] {$a$} (DONE);

\path[->, bend right = 30] (T0) edge node[left] {$c$} (S4);
\path[->, bend right = 60] (Tk) edge node[left] {$c$} (S4);



\end{tikzpicture}
\end{center}
\vspace{-5mm}
\caption{ The model $M_m$ used for the benchmarking. If $m{=0}$ then there is no states $t_i$
and $q{=}1$. If $m{>}0$ then we have states $t_0 ... t_{m{-}1}$; $p {=} 0.3/m$ and $q = 0.7$.} \label{fig.Mk}
\end{figure}
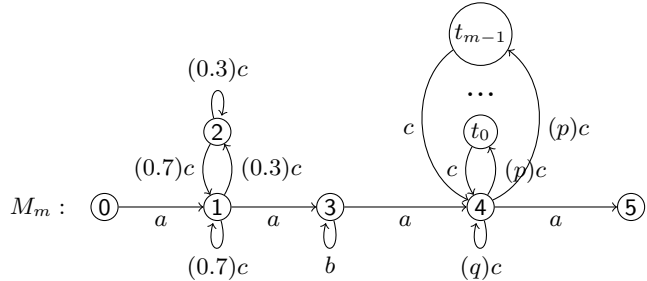

\begin{figure}[t]
\begin{center}
\begin{multicols}{2}
\scriptsize
\begin{tabular}{|c|r|r|r|c|} \hline
  & $|tc|$ & $\#nodes$ & $\#paths$ & $\#nodes^3$ \\ \hline
$E(5,0)$ & 20 & 26 & 16 & 103(4) \\ \hline
$E(6,0)$ & 23 & 30 & 32 & 144(5) \\ \hline
$E(7,0)$ & 26 & 34 & 64 & 223(7) \\ \hline
$E(8,0)$ & 29 & 38 & 128 & 381(10) \\ \hline
$E(9,0)$ & 32 & 42 & 256 & 422(10) \\ \hline
$E(10,0)$ & 35 & 46 & 512 & 501(11) \\ \hline
$E(11,0)$ & 38 & 50 & 1024 & 659(13) \\ \hline
$E(12,0)$ & 41 & 54 & 2048 & 700(13) \\ \hline
$E(5,2)$ & 20 & 34 & 336 & 185(5) \\ \hline
$E(6,2)$ & 23 & 40 & 1376 & 306(8) \\ \hline
$E(7,2)$ & 26 & 46 & 5440 & 435(9) \\ \hline
$E(8,2)$ & 29 & 52 & 21888 & 695(13) \\ \hline
$E(9,2)$ & 32 & 58 & 87296 & 944(16) \\ \hline
$E(10,2)$ & 35 & 64 & 349696 & 1073(17) \\ \hline
$E(11,2)$ & 38 & 70 & 1397760 & 1333(19) \\ \hline
$E(12,2)$ & 41 & 76 & 5593088 & 1582(21) \\ \hline
$E(5,8)$ & 20 & 58 & 3600 & 863(15) \\ \hline
$E(6,8)$ & 23 & 70 & 29984 & 2760(39) \\ \hline
$E(7,8)$ & 26 & 82 & 175168 & 4287(52) \\ \hline
$E(8,8)$ & 29 & 94 & 1309824 & 8261(88) \\ \hline
$E(9,8)$ & 32 & 106 & 8225024 & 23726(224) \\ \hline
\end{tabular}

\begin{tabular}{|c|c|c|c|c|} \hline
  & $f_1$ & $f_2$ & $f_3$ & $f_4$\\ \hline
E(5,0) & 0.001 & 0.002 & 0.001 & 0.002\\ \hline
E(6,0) & 0.001 & 0.002 & 0.001 & 0.002\\ \hline
E(7,0) & 0.001 & 0.003 & 0.001 & 0.003\\ \hline
E(8,0) & 0.001 & 0.004 & 0.001 & 0.005\\ \hline
E(9,0) & 0.001 & 0.005 & 0.002 & 0.006\\ \hline
E(10,0) & 0.001 & 0.006 & 0.003 & 0.008\\ \hline
E(11,0) & 0.001 & 0.008 & 0.004 & 0.012\\ \hline
E(12,0) & 0.001 & 0.008 & 0.009 & 0.024\\ \hline
E(5,2) & 0.001 & 0.002 & 0.002 & 0.004\\ \hline
E(6,2) & 0.001 & 0.004 & 0.002 & 0.01\\ \hline
E(7,2) & 0.001 & 0.005 & 0.003 & 0.039\\ \hline
E(8,2) & 0.001 & 0.01 & 0.005 & 0.138\\ \hline
E(9,2) & 0.001 & 0.014 & 0.01 & 0.44\\ \hline
E(10,2) & 0.001 & 0.012 & 0.019 & 1.09\\ \hline
E(11,2) & 0.001 & 0.018 & 0.041 & 3.13\\ \hline
E(12,2) & 0.001 & 0.023 & 0.091 & 10.68\\ \hline
E(5,8) & 0.001 & 0.011 & 0.006 & 0.032\\ \hline
E(6,8) & 0.001 & 0.04 & 0.034 & 0.279\\ \hline
E(7,8) & 0.001 & 0.076 & 0.073 & 1.38\\ \hline
E(8,8) & 0.002 & 0.154 & 0.266 & 12.04\\ \hline
E(9,8) & 0.002 & 0.46 & 0.539 & 219\\ \hline
\end{tabular}
\end{multicols}
\end{center}
    \vspace{-5mm}
    \caption{{\bf Left:} the execution models used in the benchmark. $\#nodes$ and $\#paths$ 
    are the number of nodes and full paths (executions) in the corresponding execution model; 
    $\#nodes^3$ is the number of nodes in the resulting 3-word expansion model. The number
    between brackets is $\#nodes^3/\#nodes$.
    {\bf Right:} the run time (seconds) of our coverage calculation algorithm on different execution models and  
    coverage goals.
    } \label{fig.experiment.emodels}
\end{figure}

The number of possible executions in the execution models correspond to their degree of non-determinism.
The test case $tc^i$ has been designed as such that increasing $i$ exponentially
increases the non-determinism of the corresponding execution model (we can see this
in Figure \ref{fig.experiment.emodels} by comparing $\#paths$ with the $i$
index of the corresponding $E(i,m)$).

All the models used ($M_0$, $M_2$, and $M_8$) are non-deterministic: $M_0$ is the
least non-deterministic one whereas $M_8$ is very non-deterministic. This is reflected
in the number of possible executions in their corresponding execution models, with $E(i,8)$
having far more possible executions than $E(i,0)$.

The following four coverage goals are used:
\[ \scriptsize
  \begin{array}{|l|c|c|} \hline
      \mbox{goal} & \mbox{type} & \mbox{word expansion} \\ \hline
      f_1 : \WORD{2} ; \WORD{t_0} & \mbox{simple sentence} & \mbox{no} \\ \hline
      f_2 : \WORD{1,1,1} ; \WORD{4,4,4} & \mbox{non-simple sentence} & \mbox{3-word} \\ \hline
      f_3 : \; ^{1}{\geq}_8 & \mbox{aggregate} & \mbox{no} \\ \hline
      f_4 : \; ^{3}{\geq}_8 & \mbox{aggregate} & \mbox{3-word} \\ \hline
  \end{array}
\]
We let our algorithm calculates the coverage of each of the above goals on 
the execution models $E(5,0)...E(9,8)$ and measure the time it takes to finish
the calculation. For the merging policy, $n$ is set to 1 when the goal does
not need word expansion, and else it is set to be equal to the expansion
parameter. 
The experiment is run on a Macbook Pro with 2,7 GHz Intel i5 and
8 GB RAM.
Fig. \ref{fig.experiment.emodels} (right) shows the results.
For example, we can see that $f_1$ can be calculated in just a few milli seconds, even on
$E(12,m)$ and $E(i,8)$. 
In contrast, brute force calculation using Def. \ref{def.tc.query.prob}
on e.g. $E(11,2), E(12,2), E(8,8)$, and $E(9,8)$ would be very expensive, because it has to quantify
over more than a million paths in each of these models.

\begin{figure}[t]
  \begin{center}
  \begin{tabular}{ll}
  \scalebox{0.3}[0.4]{\includegraphics{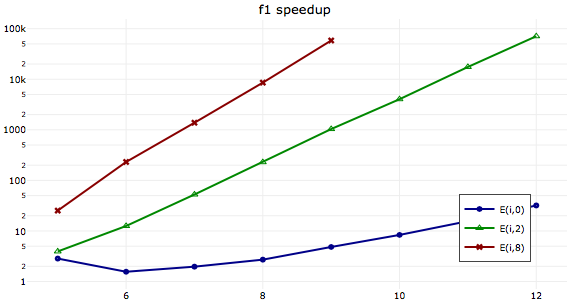}} &
  \scalebox{0.3}[0.4]{\includegraphics{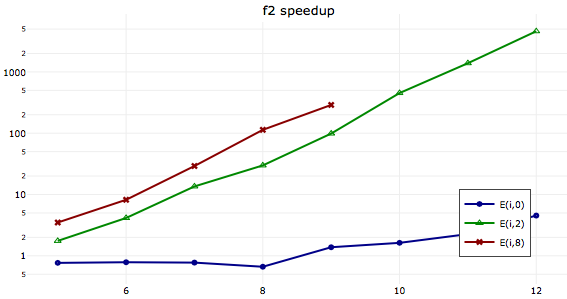}} 
  \\
  %
  \scalebox{0.3}[0.4]{\includegraphics{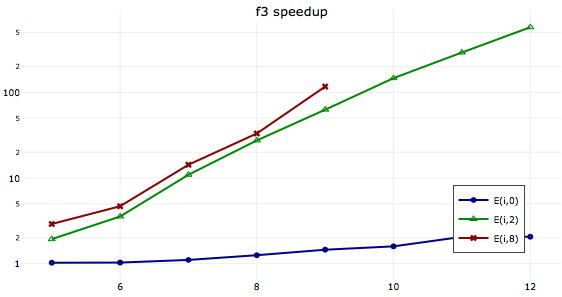}} &
  \scalebox{0.3}[0.4]{\includegraphics{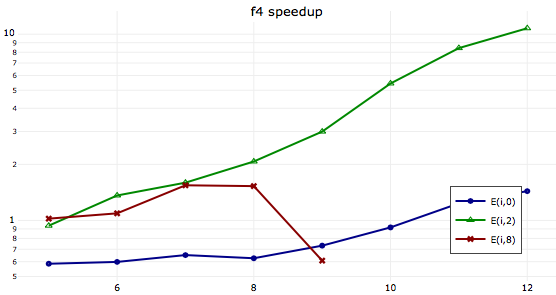}} \\
  \end{tabular}
  \end{center}
  \vspace{-5mm}
  \caption{ The graphs show our algorithm's speedup with respect to the brute force calculation on
  four different goals: $f_1$ (top left), $f_2$ (top right), $f_3$ (bottom left), and $f_4$ (bottom right).
  $f_1$ and $f_2$ are non-aggregate, whereas $f_3$ and $f_4$ are aggregate goals. Calculating
  $f_1$ and $f_3$ does not use word expansion, whereas $f_2$ and $f_4$ require 3-word expansion.
  Each graph shows the speedup with respect to three families of execution models: $E(i,0)$, $E(i,2)$, and $E(i,8)$.
  These models have increasing degree of non-determinism, with models from $E(i,8)$  being the most
  non-deterministic ones compared to the models from other families (with the same $i$). The horizontal axes represent
  the $i$ parameter, which linearly influences the length of the used test case. The vertical axes show
  the speedup in the {\bf logarithmic} scale.
   }
  \label{fig.speedup}
\end{figure}

Fig.\ref{fig.speedup}
shows the speedup of our algorithm with respect to the brute force calculation ---note that
the graphs are set in logarithmic scale. We can see that in almost
all cases the speedup grows exponentially with respect to the length of the test case,
although the growth rate is different in different situations. We can notice that the speed up on
$E(i,0)$ is much lower (though we still have speedup, except for $f_4$ which we will discuss below). 
This is because $E(i,0)$'s are not too non-deterministic. They all 
induce less than 2100 possible executions. The brute force approach can easily handle such volume.
Despite the low speedup, on all $E(i,0)$'s our algorithm can do the task in just few milli seconds 
(1 - 24 ms).

The calculation of $f_1$ is very fast (less than 2 ms). This is expected, because $f_1$ is a simple sentence. 
The calculation of $f_2$, on the other hand, which is a non-simple sentence,
must be executed on the corresponding $3$-word expanded execution model, which can be much larger
than the original execution model. E.g. $E(9,8)^3$ is over 200 times larger (in the number of nodes)
than $E(9,8)$. Despite this we see the algorithm performs pretty well on $f_2$.

$f_3$ and $f_4$ are both aggregate goals. The calculation of $f_3$ is not problematical, however
we see that $f_4$ becomes expensive on the models $E(12,2), E(8,8)$, and $E(9,8)$ (see Fig. \ref{fig.experiment.emodels} right). In fact, on $E(9,8)$ the calculation of $f_4$ is even worse than brute force (the dip in the red line
in Fig. \ref{fig.speedup}). Recall that $f_4 = \; ^{3}{\geq}_8$; so, calculating its coverage requires 
us to sum over different sets of words of size 3 that the different executions can generate.
$E(12,2), E(8,8)$, and $E(9,8)$ are large (over 70 states) and highly non-deterministic. 
Inevitably, they generate a lot of words of size 3, and therefore the number of possible sets of these words 
explodes. E.g. on $E(8,8)$ and $E(9,8)$ our algorithm ends up with about 1.2M an 6.7M sets of words
to sum over. In contrast, the number of full paths in these models are about respectively 1.3M and 8.2M.
At this ratio, there is not much to gain with respect to the brute force approach that simply sums over
all full paths, whereas our algorithm also has to deal with the overhead of book keeping and merging.
Hypothetically, if we always merge, the number of final sets 
of words can be reduced to respectively about 500K and 2M, so summing over them would be faster. We should not do this though, because merging is expensive, but the numbers do suggest that there is room for improvement
if one can figure out how to merge more smartly.




\section{Related Work} \label{sec.relatedwork}

\vspace{-3mm}
To the best of our knowledge the concept of probabilistic coverage has not been well addressed
in the literature on non-deterministic MBT, or even in the literature on probabilistic
automata. A paper by Zu, Hall, and May \cite{zhu1997software} that provides a comprehensive
discussion on various coverage criteria does not mention the concept either.
This is a bit surprising since coverage is a concept that is quite central in
software testing. We do find its mentioning in literature on statistical testing,
e.g. \cite{denise2004generic,whittaker1994markov}.
In \cite{whittaker1994markov} Whittaker and Thomason discussed the use of 
Markov chains to encode probabilistic behavioral models. The probabilities are used to model 
the usage pattern of the SUT. This allows us
to generate test sequences whose distribution follows the usage pattern (so-called 'statistical testing'). 
Techniques from Markov chain are then used to predict properties of the test sequences if we are to generate them in this way, e.g. 
the probability to obtain a certain level of node or edge coverage, or conversely the 
expected number of test runs needed to get that level of coverage.
In contrast, in our work probabilities are used to model SUT's non-determinism, rather than its usage pattern. We do not concern
ourselves with how the tester generates the test sequences, and focuses purely on the
calculation of coverage under the SUT's non-determinism. Our coverage goal expressions are
more general than \cite{whittaker1994markov} by allowing words of arbitrary length (rather than
just words of length one or two, which would represent state and respectively edge coverage),
clauses, and sentences to be specified as coverage goals. Coverage calculation in both \cite{denise2004generic,whittaker1994markov}
basically comes down to the brute force calculation in Def. \ref{def.tc.query.prob}.

Our algorithm to calculate the coverage of simple sentences has some similarity with the 
probabilistic model checking algorithm for Probabilistic Computation Tree Logic (PCTL) \cite{hansson1994logic,kwiatkowska2007stochastic}. 
Although given a formula $f$ a model checking
algorithm tries to decide whether or not $f$ is valid on the given behavior model,
the underlying probabilistic algorithm also labels for every state in the model with the probability
that any execution that starts from that state would satisfy $f$. 
Since we only need to calculate over execution models, which are acyclic,
there is no need to do a fix point iteration as in \cite{kwiatkowska2007stochastic}.
From this perspective, our algorithm can be seen as an instance of \cite{kwiatkowska2007stochastic}.
However we also add $k$-word expansion. In addition to simplifying the algorithm when dealing
with non-simple sentences, the expansion also serves as a form of memoisation (we do not have to keep
calculating the probability for a state $u$ to lead to a word $w$). 
In particular the calculation of aggregate coverage goals benefits from this memoisation.
\HIDE{
Whereas our algorithm
performs the labelling directly on the graph representing the given execution model,
the algorithm in \cite{kwiatkowska2007stochastic} uses a matrix to represent such a model
and does the labelling through matrix multiplication. In one hand this is wasteful if the
matrix is sparse, on the other hand matrix multiplication can also be parallelized. 
}
Though, the
biggest difference between our approach with a model checking algorithm is that the latter does not
deal with aggregate properties (there is no concept of aggregate formulas in PCTL). 
Our contribution can also be seen as opening a way to extend a probabilistic model checking algorithm to
calculate such properties. We believe it is also possible to generalize over 
the aggregation so that the same algorithm can be used to aggregate arbitrary state
attributes that admit some aggregation operator (e.g. the cost of staying in various states,
which can be aggregated with the '+' operator).

In this paper we have focused on coverage analyses. There are other analyses that are useful to mention.
In this paper we abstract away from the data that may have been exchanged during the interactions
with the SUT. In practice many systems do exchange data. In this situation we
may also want to do data-related analyses as well. E.g. the work by Prasetya \cite{prasetya2018temporal}
discussed the use of an extended LTL to query temporal
relations between the data exchanged through the test sequences in a test suite.
This is useful e.g. to find test sequences of a specific property, or to check if
a certain temporal scenario has been covered. The setup is non-probabilistic though
(a query can only tell whether a temporal property holds or not), so an extension
would be needed if we are interested in probabilistic judgement. 
Another example of analyses is risk analyses as in the work by Stoelinga and Timmer \cite{stoelinga2009interpreting}.
When testing a non-deterministic system, we need to keep in mind 
that although executing a test suite may report no error, there might still be lurking errors
that were not triggered due to internal non-determinism.
Stoelinga and Timmer propose to annotate each transition in a model
with the estimated probability that it is incorrectly implemented and the entailed cost 
if the incorrect behavior emerges\footnote{We gloss over the complication that the transition
might be in a cycle. A test case may thus exercise it multiple times. Each time, exercising it successfully would
arguably decrease the probability that it still hides some hidden erroneous behavior.
This requires a more elaborate treatment, see  \cite{stoelinga2009interpreting} for more details.
}. This then allows us to calculate the probability that a successful execution of a test suite
still hides errors, and the expected cost (risk) of these hidden errors. 

\HIDE{
offer an interesting perspective on the concept of probabilistic coverage.
In this work, the SUT is modeled as a function whose input and output domains are divided into a finite number of equivalence
classes. So, a finite domain function can be used to model it. Probabilities are introduced to model
the likelihood that the behavior of each input domain is incorrectly implemented by the SUT.
Since testing is incomplete, when a test suite $\Gamma$ reports no error note that it does not necessarily imply that the SUT is thus really
bug free. Stoelinga and Timmer defines coverage as, essentially,
the number of input partitions covered by $\Gamma$, multiplied by the probability that all these partitions were indeed implemented 
correctly (in other words, it factors in the probability that $\Gamma$ does not 
report any false negative into the usual structure-based coverage).
}

\section{Conclusion} \label{sec.concl}

\vspace{-3mm}
We have presented a concept of probabilistic coverage that is useful to express
the coverage of a test suite in model-based testing when the used model is
non-deterministic, but has been annotated with estimation on the probability
of each non-deterministic choice. Both aggregate and non-aggregate
coverage goals can be expressed, and we have presented an algorithm to
efficiently calculate the probabilistic coverage of such goals. Quite sophisticated
coverage goals can be expressed, e.g. sequence (words) coverage and sequence of sequences (sentences) 
coverage. We have shown that in most cases the algorithm is very efficient. A challenge still lies on
calculating aggregate $k$-wise test goals on test cases that repeatedly trigger highly non-deterministic
parts of the model. Such a situation bounds to generate combinatoric explosion on the
possible combinations of words that need to be taken into account. Beyond a certain point,
the explosion becomes too much for the merging policy used in our algorithm to handle.
Analyses on the data obtained from our benchmarking suggests that in theory there is indeed room
for improvement, though it is not yet clear what the best course to proceed. 
This is left for future work.

\bibliographystyle{splncs04}
\bibliography{thebibs}

\end{document}